\documentclass[aps,pra,singlecolumn,superscriptaddress,amsmath,amssymb,eqsecnum,nofootinbib, reprint]{revtex4-2}
\usepackage{graphicx}
\usepackage{amsthm}
\usepackage{latexsym}
\usepackage{graphics,epsfig}
\allowdisplaybreaks
\usepackage{enumitem}
\usepackage{braket}
\usepackage{cleveref}
\usepackage{graphicx}
\usepackage{subcaption}
\usepackage{caption}
\def\XXint#1#2#3{{\setbox0=\hbox{$#1{#2#3}{\int}$}
\vcenter{\hbox{$#2#3$}}\kern-.5\wd0}}

\newtheorem{proposition}{Proposition}

\theoremstyle{remark}

\def\({\left(}
\def\){\right)}
\def\[{\left[}
\def\]{\right]}
\def\{{\left\lbrace}
\def\}{\right\rbrace}
\def\<{\langle}
\def\>{\rangle}
\def\bpn{\begin{proposition}}
\def\epn{\end{proposition}}
\def\bpf{\begin{proof}}
\def\epf{\end{proof}}

\newcommand{\bb}[1]{\mathbb{#1}}

\renewcommand{\o}[1]{\overline{#1}}

\newcommand{\SU}{\mathrm{SU}}

\newcommand{\SO}{\mathrm{SO}}
\newcommand{\SL}{\mathrm{SL}}
\newcommand{\He}[1]{\mathrm{He}({#1})}

\def\bea{\begin{equation*}\aligned}
\def\eea{\endaligned\end{equation*}}
\def\bean{\begin{equation}\aligned}
\def\eean{\endaligned\end{equation}}
\def\be{\begin{equation*}\aligned}
\def\ee{\endaligned\end{equation*}}

\newcommand{\abs}[1]{\left| #1 \right|}

\numberwithin{equation}{section}

\numberwithin{proposition}{section}

\begin{document}

\title{Qutrit codes within representations of SU(3)}
\author{Xzavier Herbert}
\email{xzavier.herbert@gmail.com}
\affiliation{Massachusetts Institute of Technology, Cambridge, MA 02139}

\author{Jonathan Gross}
\affiliation{Google Quantum AI, Santa Barbara, California 93117, USA}

\author{Michael Newman}
\affiliation{Google Quantum AI, Santa Barbara, California 93117, USA}
\date{\today}

\begin{abstract}
    We describe a quantum error-detecting and error-correcting code embedded within irreducible representations of SU(3).  These logical qutrits inherit the He(3) symmetries induced by the representation, while protecting against small SU(3) displacements.  We explore the general methodology for finding codes from structure-inducing representations of groups, together with symmetries inherited from finite subgroups, extending the case of spin representations of SU(2) \cite{gross_designing_2021}.
\end{abstract}

\maketitle

\section{Introduction}

Quantum error correction is essential to realizing large-scale quantum computation.
To distinguish errors from computation, one must embed the logical quantum state space in a larger physical state space.
The most widely used strategy for doing this is to encode a small number of logical systems in a large number of physical systems, such as encoding a single logical qubit in a collection of many physical qubits.

However, depending on the types of errors one wants to correct, the large physical state space need not decompose into multiple subsystems.
Gottesman-Kitaev-Preskill (GKP) codes~\cite{gottesman_encoding_2001}, cat codes~\cite{cochrane_macroscopically_1999}, binomial codes~\cite{michael_new_2016}, and other variants~\cite{chuang_bosonic_1997,grimsmo_quantum_2019} all encode logical qubits in the infinite-dimensional Hilbert space of a bosonic mode.
While multi-system codes are designed to protect against errors restricted to a few subsystems, these bosonic codes are designed to protect against dominant errors in oscillator systems, such as photon loss, dephasing, thermalization, and displacement~\cite{glancy_error_2006,albert_performance_2018,noh_quantum_2019}.
Because harmonic oscillators are ubiquitous, these codes have enjoyed a great deal of experimental attention~\cite{ofek_extending_2016,rosenblum_fault-tolerant_2018,fluhmann_encoding_2019,campagne-ibarcq_stabilized_2019,guillaud_repetition_2019,hu_quantum_2019,lescanne_exponential_2020}.

Of course, there are many monolithic state spaces other than bosonic modes.
Qubits have also been encoded in large angular-momentum state spaces~\cite{albert_robust_2020,gross_designing_2021,omanakuttan_multispin_2023,xu_clifford_2023,jain_ae_2023} and abstract finite-dimensional systems~\cite{pirandola_minimal_2008,cafaro_quantum_2012}.
In these cases, the relevant errors are given by some extra structure on the Hilbert space.
For bosonic modes, the relevant Gaussian errors correspond to a representation of $\SL(2,\mathbf{R})$ on the Hilbert space.
For angular-momentum systems, there is a natural representation of $\SU(2)$ or $\SO(3)$, and more abstract systems are generally associated with a representation of a discrete group, such as a cyclic group~\cite{pirandola_minimal_2008} or finite Heisenberg group~\cite{cafaro_quantum_2012}.

The present work is a natural extension of the approach followed in~\cite{gross_designing_2021}, where we take the physical state space to be an irreducible representation of $\SU(3)$.
In addition to defining relevant errors, the structure-inducing representation also defines the simplest controls available over the physical system.
We therefore desire to find codes that not only correct small $\SU(3)$ displacements, but also allow certain logical operations to be performed using large $\SU(3)$ displacements.
This is analogous to transversal gates in multi-system codes, which allow logical operations when acting on many of the subsystems, as well as the GKP code, which allows logical operations via large displacements.
The authors of~\cite{kubischta_family_2023,kubischta_not-so-secret_2023} have extensively studied the relationship between these different notions of transversality.
In the case of $\SU(2)$ codes, narrowing the search by restricting to highly symmetric codespaces - namely those endowed with ``natural'' logical operators - proved fruitful for efficiently protecting against errors, and we find the same true for $\SU(3)$.
In this work, we focus on enforcing He(3) symmetry, endowing our codes with the analogue of logical Pauli operators.

The paper is organized as follows.
We begin with a brief introduction in Section~\ref{background}, including the Knill-Laflamme conditions for quantum error-correction and the relevant representation theory of He(3) $\subseteq$ SU(3).
In Section~\ref{reduction}, we discuss using the imposed He(3) symmetry to simplify the detection and correction conditions. 
In Section~\ref{detecting}, we present a $15$-dimensional quantum error-detecting code which supports logical Pauli operators.  
We also use this simpler example to build intuition for tools that will assist our search for a quantum error-correcting code.
Finally, in Section~\ref{correcting} we present a much larger quantum error-correcting code along with a proof of  correctability against small $\SU(3)$ displacements.
We conclude in Section \ref{conclusions} with thoughts on how to find smaller codes, codes with more symmetries, and plausible physical instantiations of such codes.  For convenience, Python code verifying correctability may be found at \cite{python_package}.

\section{Background}
\label{background}

\subsection{Error correction conditions}

We begin with the group SU(3), generalizing the ``spin code'' case of SU(2).  
Our aim is to find a special subspace of a monolithic state space $\mathcal{H}$ presented as an irreducible representation of SU(3).  
The representation imbues $\mathcal{H}$ with distinguished error operators - those generated by $\mathfrak{su}$(3).  
Correcting such error operators are relevant e.g. to number-preserving operations on collections of harmonic oscillators \cite{schwinger_angular_1952}.  

The Knill-Laflamme quantum error correction conditions determine whether particular error can be corrected.  
Given a code spanned by an orthonormal basis $\{\ket{\o i}\}$, we say the code detects a set of errors $\{E_a\}$ if $$\bra{\o i} E_a \ket{\o j} = C_{a}\delta_{ij}$$ and corrects them if $$ \bra{\o i} E_b^\dagger E_a \ket{\o j} = C_{ab} \delta_{ij}.$$
It suffices to satisfy these conditions on a basis for a set of errors.  For $\mathfrak{su}(3)$, we consider representatives for the basis
\begin{equation*}
\begin{array}{cc}
\lambda_1 = \(\begin{array}{ccc}
     0& 1&0 \\
     1& 0&0 \\
     0&0&0
\end{array}\) &
\lambda_2 = \(\begin{array}{ccc}
     0& 0&1 \\
     0& 0&0 \\
     1&0&0
\end{array}\) \\
\lambda_3 = \(\begin{array}{ccc}
     0& 0&0 \\
     0& 0&1 \\
     0&1&0
\end{array}\) &
\lambda_4 = \(\begin{array}{ccc}
     0& -i&0 \\
     i& 0&0 \\
     0&0&0
\end{array}\) \\
\lambda_5 = \(\begin{array}{ccc}
     0& 0&-i \\
     0& 0&0 \\
     i&0&0
\end{array}\) &
\lambda_6 = \(\begin{array}{ccc}
     0& 0&0 \\
     0& 0&-i \\
     0&i&0
\end{array}\) \\
H_1 = \(\begin{array}{ccc}
     1& 0&0 \\
     0& -1&0 \\
     0&0&0
\end{array}\) &
H_2 = \(\begin{array}{ccc}
     0& 0&0 \\
     0& 1&0 \\
     0&0& -1
\end{array}\)
\end{array}.
\end{equation*}

\subsection{Irreducible representations of SU(3)}

We briefly review the representation theory of SU(3) - see \cite{hall2000elementary} for further details.  The irreducible representations of SU(3) are classified in terms of their highest weights, which are determined by vectors in the associated complexified Lie algebra $\mathfrak{su}(3)_{\mathbb{C}} \cong \mathfrak{sl}(3; \bb C)$. 
For any representation $(\pi, V)$ of $\mathfrak{sl}(3; \bb C)$, a weight for $\pi$ is an ordered pair $\mu = (\mu_1, \mu_2) \in \mathbb{Z}^2_{\geq 0}$ such that there is some nonzero $v \in V$ such that 
\be
\pi(H_1)v &= \mu_1 v\\
\pi(H_2)v &= \mu_2 v.
\ee
A root is defined similarly as a nontrivial weight of the adjoint representation, i.e. an ordered pair $\alpha = (\alpha_1, \alpha_2) \in \bb C^2$ such that $(\alpha_1, \alpha_2) \neq 0$ and there exists $Z \in \mathfrak{sl}(3 ; \bb C)$ such that
\be
\[H_1, Z\] & = \alpha_1 Z\\
\[H_2, Z\] &= \alpha_2 Z.
\ee
The ordering on weights is defined using the respective roots of the raising operators $\ket{1}\bra{0}$ and $\ket{2}\bra{1}$,
\be
\alpha &= (2, -1)\\
\beta &= (-1,2).
\ee
The partial ordering is then defined as $\mu_1 \succeq \mu_2$ if there exist $a,b \geq 0$ such that
\be
\mu_1 - \mu_2 = a\alpha + b\beta.
\ee
Each irreducible representation of SU(3) has a unique highest weight under this ordering, and so we refer to the irreducible representations using these highest weights $(p,q)$.

\subsection{$(p,0)$ representations}

The standard prescription for constructing $(p,q)$ representations is to identify them within the combined $p$- and $q$-fold tensor product of the $(1,0)$ and $(0,1)$ representations, respectively.  
The $(1,0)$ representation corresponds to the standard representation, with highest weight vector $\ket{0}$, and $(0,1)$ its dual, with highest weight vector $\ket{2}$.
One can repeatedly apply sequences of lowering operators $Y_1$ and $Y_2$ to the highest $(p,q)$-weight vector $\ket{0}^{\otimes p} \otimes \ket{2}^{\otimes q}$ (under the product rule action by $\mathfrak{sl}(3;\mathbb{C})$) to explicitly construct the $(p,q)$ representation.
In general, it has dimension
\be
\frac{1}{2}(p + 1)(q+1)(p+q+2).
\ee

To assist our search, we focus on $(p,0)$ representations of $SU(3)$, which follow a simple pattern.
The $(p,0)$ representations are generated by symmetric vectors, taking the form
\be
\[a_1 a_2 \dots a_p\] := \sum_{\sigma \in \operatorname{orb}_{S_3} \[a_1a_2\dots a_p\]} \sigma.
\ee
Note that (as expected) the number of symmetric vectors $\binom{p+2}{2}$ matches the dimension formula.

\subsection{Representations of $He(3)$}

Of similar importance to protecting encoded information is to be able to deliberately manipulate it.
Consequently, we will focus on quantum codes which are endowed with logical He(3) actions inherited from the SU(3) representation.
Importantly, He(3) is a finite group, so we can distinguish different logical operations from unintended infinitesimal transformations \cite{eastin_restrictions_2009}.
Being a finite group, $\He{3}$ has finitely many irreducible representations. These representations are given in Table~\ref{character}.
\begin{table*}[ht]\centering
\begin{tabular}{c|ccccccccccc}
     
      & $I$ & $\omega I$ & $\omega^2 I$ & $X$ & $X^2$ & $Z$ & $Z^2$ & $XZ$ & $X^2Z$ & $XZ^2$ & $X^2Z^2$ \\
     \hline
     $\chi_{std}$ & 3 & $3\omega$ & $3\omega^2$ & 0 &$\cdots$ \\
     $\chi_{std^{op}}$ & 3 & $3\omega^2$ & $3\omega$ & 0 & $\cdots$\\
     $\chi_{triv} = \chi_{\omega, X, Z_1}$ & 1 & 1& $\cdots$\\
     $\chi_{\omega, X_1}$ & 1&1&1 &1&1 & $\omega$ & $\omega^2$ & $\omega$ & $\omega$ & $\omega^2$ & $\omega^2$\\
     $\chi_{\omega, X_2}$ & 1&1&1 &1&1 & $\omega^2$ & $\omega$ & $\omega^2$ & $\omega^2$ & $\omega$ & $\omega$\\
     $\chi_{\omega, Z_1}$ & 1 & 1 & 1  & $\omega$ & $\omega^2$ & 1 & 1  & $\omega$ & $\omega^2$ & $\omega$ & $\omega^2$\\
     $\chi_{\omega, Z_2}$ & 1 & 1 & 1  & $\omega^2$ & $\omega$ & 1 & 1  & $\omega^2$ & $\omega$ & $\omega^2$ & $\omega$\\
     $\chi_{\omega, XZ_1}$ & 1 & 1 & 1  & $\omega$ & $\omega^2$ & $\omega^2$ & $\omega$  & 1 & $\omega$ & $\omega^2$ & 1\\
     $\chi_{\omega, XZ_2}$ & 1 & 1 & 1  & $\omega^2$ & $\omega$ & $\omega$ & $\omega^2$  & 1 & $\omega^2$ & $\omega$ & 1\\
     $\chi_{\omega, X^2Z_1}$ & 1 & 1 & 1  & $\omega^2$ & $\omega$ & $\omega^2$ & $\omega$  & $\omega$ & 1 & $1$ & $\omega^2$ \\
     $\chi_{\omega, X^2Z_2}$ & 1 & 1 & 1  & $\omega$ & $\omega^2$ & $\omega$ & $\omega^2$  & $\omega^2$ & 1 & $1$ & $\omega$ \\
\end{tabular}
\caption{Character table of He(3).}
\label{character}
\end{table*}

\begin{figure}[!ht]
\begin{subfigure}{0.95\columnwidth}
    \includegraphics[width=0.9\columnwidth]{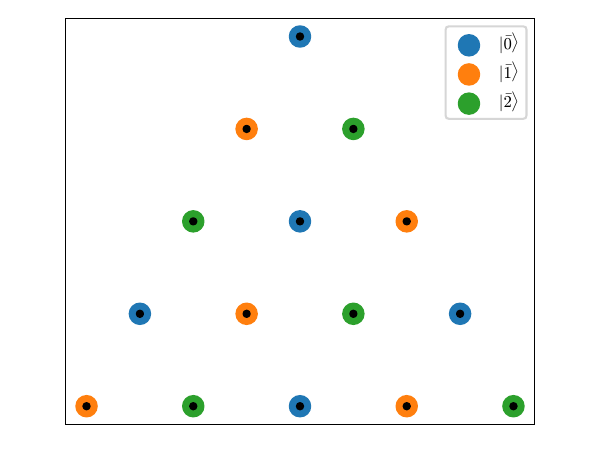}
    \subcaption{For $p\equiv1\mod3$ we can obtain standard representations of $\He{3}$, since the rotations made by $X$ permute the eigenstates of the $Z$ representative in the ordinary direction.}
    \label{fig:p4-supports}
\end{subfigure}

\begin{subfigure}{0.95\columnwidth}
    \includegraphics[width=0.9\columnwidth]{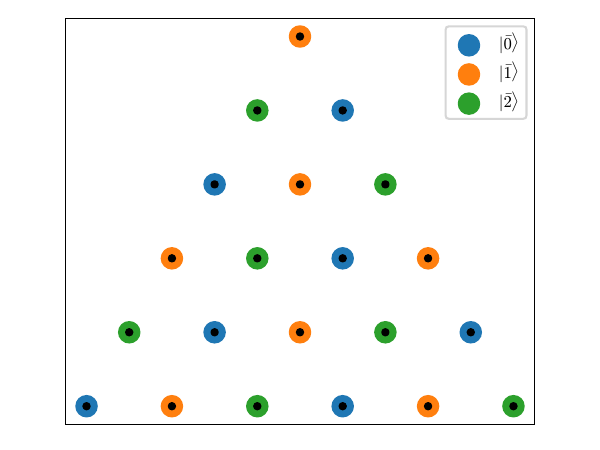}
    \caption{For $p\equiv2\mod3$ we can obtain dual representations of $\He{3}$, since the rotations made by $X$ permute the eigenstates of the $Z$ representative in the opposite direction.}
    \label{fig:p5-supports}
\end{subfigure}

\begin{subfigure}{0.95\columnwidth}
    \includegraphics[width=0.9\columnwidth]{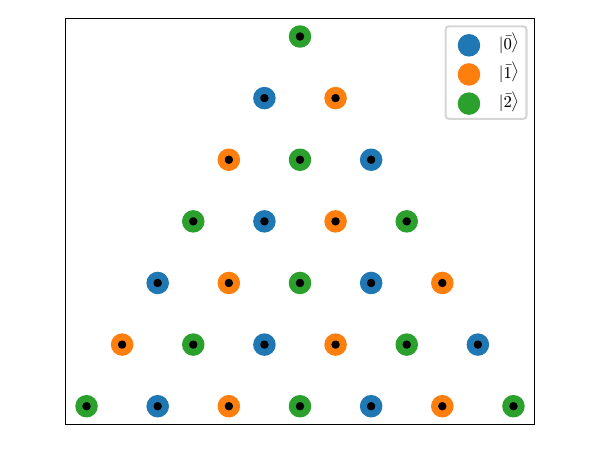}
    \caption{For $p\equiv0\mod3$ we can't obtain faithful representations of $\He{3}$, since the rotations made by $X$ do not permute all three of the $Z$-eigenstates.}
    \label{fig:p6-supports}
\end{subfigure}
\caption{Candidates for faithful representations of He(3) within $(p,0)$ representations of SU(3).}
\label{candidates}
\end{figure}

To obtain a faithful representation of $\He{3}$, the representative of $Z$ must apply the appropriate phases on the logical states.
Since $\omega Z=\exp\big[-i2\pi(H_1+H_2)/3\big]$, the weight vectors are also eigenvectors of $Z$.
This means we can make diagrams of the weight vectors forming the supports of various logical states, as in Figure~\ref{candidates}.
These diagrams also let us easily see what happens when we apply the $X$ operator - it simply rotates the triangle counter clockwise.
From this, we observe that faithful representations of He(3) are only possible when $p\not\equiv0\mod3$.  
In fact, we focus on $p\equiv1\mod3$ to obtain standard He(3) actions.  In particular, the $\chi_{(p,0)}$ irreducible representations of SU(3) restrict to $\chi_{std}^{\oplus\frac{1}{6}(p+1)(p+2)}$ reducible representations of He(3).

\section{Constraint reduction}
\label{reduction}
Not only does embedding within faithful He(3) representations imbue the codespace with desirable logical operators - it places structural constraints on the codespace.
This helps simplify the system of equations required to satisfy the Knill-Laflamme conditions. 

Focusing on error detection for example, suppose we find a codespace on which He(3) is represented faithfully and satisfies the identical expectation condition
\be
\bra{\o 0} \lambda_1 \ket{\o 0} = \bra{\o 1} \lambda_1 \ket{\o 1} = \bra{\o 2} \lambda_1 \ket{\o 2} = C_{\lambda_1}.
\ee
Then, noting that
\be
\begin{array}{cc}
X^\dagger \lambda_1 X = \lambda_2, & (X^2)^\dagger \lambda_1 X^2 = \lambda_3,
\end{array}
\ee
we simultaneously obtain
\be
C_{\lambda_1} &=  \bra{\o 0}X^\dagger \lambda_1 X \ket{\o 0} = \bra{\o 0} \lambda_2 \ket{0}  = C_{\lambda_2}\\
C_{\lambda_1} &= \bra{\o 0}(X^2)^\dagger \lambda_1 X^2 \ket{\o 0} = \bra{\o 0} \lambda_3 \ket{\o 0} = C_{\lambda_3}.
\ee
because of the logical action of $X$ on the logical qutrits. Furthermore,
\be
C_{\lambda_1} &= \bra{\o 2} \lambda_1 \ket{\o 2} = \bra{\o 1} X^\dagger \lambda_1 X \ket{\o 1} = \bra{\o 1 } \lambda_2 \ket{\o 1}\\
C_{\lambda_1} &= \bra{\o 0} \lambda_1 \ket{\o 0} = \bra{\o 2}X^\dagger \lambda_1 X \ket{\o 2} = \bra{\o 2} \lambda_2 \ket{\o 2}\\
C_{\lambda_1} & = \bra{\o 2} \lambda_2 \ket{\o 2} = \bra{\o 1}X^\dagger \lambda_2 X \ket{\o 1} = \bra{\o 1} \lambda_3 \ket{\o 1}\\
C_{\lambda_1} &= \bra{\o 0} \lambda_2 \ket{\o 0} = \bra{\o 2} X^\dagger \lambda_2 X \ket{\o 2} = \bra{\o 2} \lambda_3 \ket{\o 2}.
\ee
Using a similar argument, we can show that if $\bra{\o i} \lambda_1 \ket{\o j} = 0$, then this condition will hold for $\lambda_2$ and $\lambda_3$ as well.  In summary, if we have a code that detects some error $\lambda_i$ for $1 \leq i \leq 3$,
then it detects any error $\lambda_j$ for $1 \leq j \leq 3$.

Next, we observe that a code that detects $\lambda_1$ errors must also detect $\lambda_4$ errors. First, we find that
\be
Z^\dagger \lambda_1 Z = -\frac{1}{2} \lambda_1 - \frac{\sqrt 3}{2} \lambda_4.
\ee
So, we can replace
\be
C_{\lambda_1} = \bra{\o 0}Z^\dagger \lambda_1 Z \ket{\o 0}  = \bra{\o 0}(-\frac{1}{2}\lambda_1 - \frac{\sqrt 3}{2}\lambda_4) \ket{\o 0},
\ee
noting that $\ket{\o 0}$ must be a $+1$-eigenvector of $Z$. Simplifying, we see
\be
\bra{\o 0} \lambda_4 \ket{\o 0} = -\sqrt{3}C_{\lambda_1}.
\ee
Next, $X^\dagger Z^\dagger \lambda_1 Z X = -\frac{1}{2} \lambda_2 - \frac{\sqrt{3}}{2} \lambda_5$ and recalling that the expectation of a codeword under $\lambda_2$ is the same as under $\lambda_1$,
\be
\bra{\o 0} \lambda_5 \ket{\o 0} = -\sqrt{3}C_{\lambda_1}.
\ee
Using a similar argument, we find that $\bra{\o 0} \lambda_6 \ket{\o 0} = -\sqrt{3}C_{\lambda_1}$. Applying the same strategy as before, we conclude that if the code detects some error $\lambda_i$, then it detects any error $\lambda_j$.

It follows similarly that a code that detects $H_i$ also detects $H_j$.
Furthermore, because codewords are eigenvectors of the $H_k$ operators,
\be
\bra{\o i} H_k \ket{\o j} = 0
\ee
for $i \not= j$. Consequently, for $H_k$ operators, we are only left to satisfy the identical expectation condition.
We will show later that this reduces down to 2 constraints. It follows that in order to find an error detecting code, there are only 8 conditions to satisfy, reduced from a naive count of 40 (noting Hermicity of the error basis). This gives a flavor of how symmetry simplifies our search, and we will observe similar reductions can be made for error correction.

\section{Finding an error-detecting code}
\label{detecting}

As we've observed, symmetries significantly reduce the set of error-detection constraints we must satisfy.
In particular, it suffices to show
\be
\bra{\o i} \lambda_1 \ket{\o j} = C_{\lambda_1}\delta_{ij} \text{ and }
\bra{\o i} H_1 \ket{\o i} = C_{H_1}.
\ee
Such a code can be found in a space as small as the $(4,0)$ representation. We start by restricting the support of $\ket{\o 0}$ to
\be
\ket{\o 0} &= \alpha \[0000\] + \beta\[1110\] + \gamma \[2220\] + \delta\[1122\] \\
&+ \epsilon\[0012\]
\ee
with $\ket{\o 1} = X \ket{\o 0}$ and $\ket{\o 2} = X \ket{\o 1}$.

We first tackle the $\lambda_1$ constraints. To do so, we compute the $\lambda_1$ action on the basis vectors, noting that $$\pi_{(4,0)} \lambda_1(v) = \(\sum_{i = 0}^3 I^{\otimes i} \otimes \lambda_1 \otimes I^{\otimes 3-i}\)(v),$$
\be
\lambda_1 \ket{\o 0} &= \alpha \[0001 \] + 4\beta \[1111\] + 2\beta\[0110\]\\ &+ \gamma\[2221\] + \delta \[0122 \] + 2\epsilon \[1022\]\\
&+ 3\epsilon \[0002\]\\
\lambda_1 \ket{\o 1} &= \alpha\[1110\] + \beta\[2220\] + 2\gamma\[0011\]\\ &+ 4\gamma\[0000\] + \delta\[2210\] + 3\epsilon\[1112\]\\ 
&+ 2\epsilon\[0012\]\\
\lambda_1 \ket{\o 2} &= \beta\[0012\] + \gamma\[0112\] + 3\delta\[0001\]\\ &+ 3\delta\[1110\] + 2\epsilon\[1122\] + 2\epsilon\[2200\].
\ee
Note that $\lambda_1\ket{\o 0}$ does not contain any $+1$-eigenvectors of $Z$. Similarly, $\lambda_1 \ket{\o 1}$ and $\lambda_1 \ket{\o 2}$ do not contain any of their respective $Z$-operator eigenstates as well. Therefore, we obtain
\be
\bra{\o 0} \lambda_1 \ket{\o 0} = C_{\lambda_1} &= 0\\
\bra{\o 1} \lambda_1 \ket{\o 1} = C_{\lambda_1} &= 0\\
\bra{\o 2} \lambda_1 \ket{\o 2} = C_{\lambda_1} &= 0.
\ee
Next, we must evaluate $\bra{\o i} \lambda_1 \ket{\o j}$, for $i \not = j$. We observe the constraints
\be
\bra{\o 0} \lambda_1 \ket{\o 1} &= 4\alpha\o \gamma +4\beta \o \alpha+ 4\gamma\o \beta + 24\abs{\epsilon}^2 =0\\
\bra{\o 0} \lambda_1 \ket{\o 2} &= 12\beta\o\delta + 12\delta\o\epsilon + 12\epsilon\o\beta = 0\\
\bra{\o 1} \lambda_1 \ket{\o 2} &= 12\delta\o{\gamma} + 12 \gamma \o{\epsilon} + 12 \epsilon \o{\delta} = 0
\ee
and note $\bra{\o i} \lambda_1 \ket{ \o j} = \bra{\o j} \lambda_1 \ket{\o i}$. Next, we compute the $H_1$ determined by the action
\be
H_1 \ket{\o 0} &= 4\alpha \[0000\] -2\beta\[1110\] \\ &+ \gamma\[2220\] -2\delta\[1122\] + \epsilon\[0012\]\\
H_1 \ket{\o 1} &= -4\alpha\[1111\] -\beta\[2221\] \\ &+ 2\gamma\[0001\] + 2\delta\[2200\] - \epsilon\[1120\]\\
H_1\ket{\o 2} &= 3\beta\[0002\] - 3\gamma\[1112\].
\ee
yielding expectations
\be
\bra{\o 0} H_1 \ket{\o 0} &= 4\abs{\alpha}^2 - 8\abs\beta^2 + 4\abs\gamma^2 \\&- 12\abs\delta^2 + 12\abs\epsilon^2\\
\bra{\o 1} H_1 \ket{\o 1} &= -4\abs\alpha^2 - 4\abs\beta^2 + 8\abs\gamma^2 \\&+ 12\abs\delta^2 - 12\abs\epsilon^2\\
\bra{\o 2}H_1\ket{\o 2} &= 12\abs\beta^2 - 12\abs\gamma^2.
\ee
Note that $\bra{\o 0} H_1 \ket{\o 0} + \bra{\o 1} H_1 \ket{\o 1} + \bra{\o 2} H_1 \ket{\o 2} = 0$. Imposing the Knill-Laflamme conditions we see that
\bea
3C_{H_1} = 0 \Rightarrow C_{H_1} = 0.
\eea
Hence, $\bra{\o 0} H_1 \ket{\o 0} = \bra{\o 1} H_1 \ket{\o 1} = \bra{\o 2} H_1 \ket{\o 2}$ must all be identically zero.  Therefore, we must have $\abs{\beta}^2 = \abs{\gamma}^2$ and the remaining constraint
\be
\bra{\o 0} H_1 \ket{\o 0} = 4\abs{\alpha}^2 - 4\abs\beta^2 - 12\abs\delta^2 + 12\abs\epsilon^2 = 0.
\ee

\begin{figure}
    \centering
    \includegraphics[width=\columnwidth]{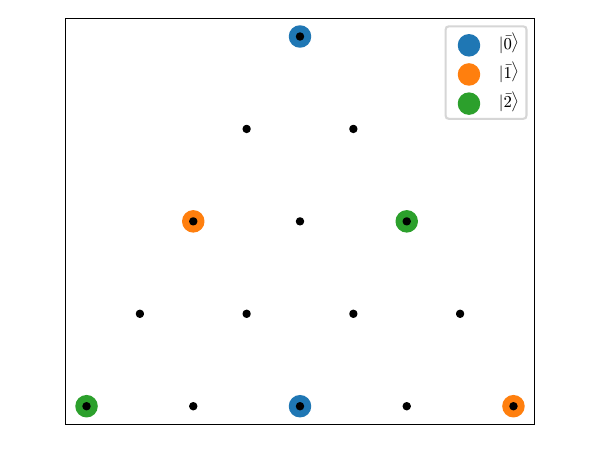}
    \caption{Visualizing the support of the error-detecting code.
    Necessary conditions for detecting the $X_j$ and $Y_j$ errors are manifest in the spacing on the triangular lattice, since these errors only move states to the adjacent weights.}
    \label{visualize_detect}
\end{figure}

Assembling the remaining constraints, we obtain the system of equations
\bea
\bra{\o 1}\lambda_1 \ket{\o 0} &= 4\alpha\o \gamma +4\beta \o \alpha+ 4\gamma\o \beta + 24\abs{\epsilon}^2 =0\\
\bra{\o 2} \lambda_1 \ket{\o 0} &= 12\beta\o\delta + 12\delta\o\epsilon + 12\epsilon\o\beta = 0\\
\bra{\o 1} \lambda_1 \ket{\o 2} &= 12\delta\o{\gamma} + 12 \gamma \o{\epsilon} + 12 \epsilon \o{\delta} = 0
\\
\bra{\o 0} H_1 \ket{\o 0} &= 4\abs{\alpha}^2 - 4\abs\beta^2 - 12\abs\delta^2 + 12\abs\epsilon^2 = 0.
\eea
To solve this system, we start by setting $\beta = \gamma = 0$, with the topmost equation forcing $\epsilon=0$. With this, it follows that $\bra{\o 1} \lambda_1 \ket{\o 2} = 0$ as desired. Consequently, we are left with a family of solutions
\be
4\abs{\alpha}^2 - 12\abs{\delta}^2 = 0 \Rightarrow \abs{\delta} = \pm \frac{1}{\sqrt 3}\abs{\alpha}
\ee
with e.g. one possible choice $\alpha = \sqrt 3$ and $\delta = 1$. The resulting codespace is given by
\be
\ket{\o 0} &= \frac{1}{3}\(\sqrt 3 \[0000\] + \[1122\]\)\\
\ket{\o 1} &= \frac{1}{3} \( \sqrt 3 \[1111\] + \[2200\] \)\\
\ket{\o 2} &= \frac{1}{3} \( \sqrt 3 \[2222\] +  \[0011\] \).
\ee
Note that the logical basis is expressed in terms of symmetric orbits, so that it is indeed normalized.
We visualize the code support in Figure~\ref{visualize_detect}, which conveys it's ability to detect bit shifts.

With the help of symmetries simplifying the error-detection conditions, a solution - defining a quantum error-detecting code - can be found in a relatively small 15-dimensional representation. 
However, the error-correction conditions introduce substantially more constraints, resulting in a blowup of the dimension of the representation.

\section{Finding an error-correcting code}
\label{correcting}

For finding an error correction code, we use various techniques to reduce the complexity of the system of equations induced by the Knill-Laflamme conditions. A key technique we employ is choosing a large enough irrep of $SU(3)$ so that the predictable action of the operators can help us choose a manageable codespace. From there, we can set certain degrees of freedom equal to zero to automatically satisfy many constraints.

\subsection{Satisfying $\lambda_i$ Constraints}

In this section, we aim to highlight strategies that will make satisfying the error detecting and correcting conditions introduced by the $\lambda_i$ operators easier. Throughout this section, we will assume that $p \equiv 1 \mod 3$. This is so that the $Z$ operator's action is equivalent its standard representation action. Define $\[s_{\ell,y, t}\]$ such that
\bea
\[ s_{\ell, y,  t}\] = \[0^{\ell} \otimes 1^y \otimes 2^t\].
\eea
We will use this notation for the rest of the paper. Note that
\bea
\[s_{\ell,y,t}\] = 0
\eea
if $\ell, y,$ or $t$ is less than 0. One convenient fact to note is that the length of $\[s_{\ell, y ,t}\]$ denoted by $\abs{\[s_{\ell, y ,t}\]}$ is given by
\bea
\abs{\[s_{\ell, y ,t}\]} = \sqrt{\frac{(\ell + y+ t)!}{\ell!y! t!}}.
\eea

Let us consider the action of $\lambda_1$ on the symmetric vector $\[s_{\ell,y, t}\]$. Letting $y \equiv t \mod 3$, from the previous section, we see that
\bea
\[ s_{\ell, y , t}\] \xrightarrow{\lambda_1}& (\ell + 1)\[s_{\ell + 1, y-1, t} \]\\
&+ (y+1)\[s_{\ell-1, y+1, t}\].
\eea
Notice that $\lambda_1$ sends a vector in the $+1$ eigenspace of the $Z$ operator to the sum of vectors in the $\omega$ and $\omega^2$ eigenspaces of $Z$. Hence,
\bea
C_{\lambda_1} = \bra{\o 0} \lambda_1 \ket{\o 0} = 0.
\eea

Because $\lambda_1$ sends $\ket{\o 0}$ to the $\omega$ and $\omega^2$ eigenspaces of the $Z$ operator, there is a possibility that $\bra{\o 1} \lambda_1 \ket{\o 0} \not= 0$ and $\bra{\o 2}\lambda_1 \ket{\o 0} \not= 0$. We can force these force these equations to be 0 by taking advantage of large $SU(3)$ irreps.

There are an unlimited number of $SU(3)$ irreps to work with, and looking in exceptionally large representations is advantageous. However, in doing so, we expose ourselves to a headache-inducing number of degrees of freedom. For instance, in Section \ref{sec: C-(37,0)} we find an error correction code in the $(37,0)$ representation. The number of degrees of freedom in this representation is $\frac{1}{2}(37 + 1)(37 + 2)\cdot \frac{1}{3} = 247$. If we considered all of the degrees of freedom, we would have a system of many equations with 247 unknowns. Using simplification techniques, we can simply set 243 of the unknowns to be zero! 

When choosing which degrees of freedom are zero, understanding the images of symmetric vectors under $\lambda_i$ gives us helpful insight. To start, we will work with the $\lambda_1$ operator. First, we will define some notation. Suppose we are given a codespace in the $(p,0)$ irrep, and let $\[s_{\ell, y ,t}\]$ be a summand of the $\ket{\o 0}$ codeword. Define the coefficient of $\[s_{\ell, y ,t}\]$ by $c_{\ell, y ,t}$ and define the sets $S_{\ket{\o i}}$ as the following
\bea
S_{\ket{\o 0}} &= \{\[s_{q, p ,r}\] \in V_{(p,0)} \mid c_{q, p ,r} \not= 0\}\\
S_{\ket{\o 1}} &= \{\[s_{q, p ,r}\] \in V_{(p,0)} \mid c_{p, r ,q} \not= 0\}\\
S_{\ket{\o 2}} &= \{\[s_{q, p ,r}\] \in V_{(p,0)} \mid c_{r, q ,p} \not= 0\},
\eea
where $V_{(p,0)}$ is the irreducible representation corresponding to the $(p,0)$ irrep. The $S_{\ket{\o i}}$ tells us what symmetric vectors appear as nonzero summands of the $\ket{\o i}$ codeword. For the following example, suppose that $\ell, y, t \geq 2$. We make the assumption that $\ell, y, t \geq 2$ for simplicity. Under the image of $\lambda_1$, $\[ s_{\ell,y, t}\]$ maps to a vector that contains $(\ell + 1)\[s_{ \ell + 1, y-1, t}\]$ and $ (y+1)\[s_{\ell - 1, y+1, t}\]$ as a summands, which are in the  $\omega^2$ and $\omega$ codespaces, respectively.

Now, we restrict our attention to $(y+1)\[s_{\ell -1, y+1, t}\]$. By definition, $\ket{\o 1}$ contains symmetric vectors that are in the $\omega$ eigenspace of the $Z$ operator. As a consequence, $\ket{\o 1} = X\ket{\o 0}$. Our aim is to find coefficient of the vector in $S_{\ket{\o 0}}$ that maps to the vector $\[s_{\ell - 1, y + 1, t}\]$ under the $X$ operator. We have the following,
\bea
X^{-1}\[s_{\ell - 1,{y+1}, t}\] = \[s_{y+1, t, \ell -1}\].
\eea
Hence, if $\[s_{y+1, t, \ell -1}\] \in S_{\ket{\o 0}}$, meaning that its coefficient $c_{y+1, t, \ell - 1} \not= 0$, then $c_{y+1, t, \ell - 1}\[s_{\ell - 1, y + 1, t}\]$ is a summand of the $\ket{\o 1}$ codeword as desired.

Next, restricting our attention to $(\ell + 1)\[s_{\ell+1, y-1 ,t}\]$, we can repeat a similar process noting that $\ket{\o 2}$ is a sum of of $\omega^2$ eigenvectors of the $Z$ operator and $\ket{\o 2} = X^2\ket{\o 0}$. We find that if $\[s_{t,\ell+1, y-1}\] \in S_{\ket{\o 0}}$, its coefficient $c_{t, \ell + 1, y-1} \not= 0$, and thus, $c_{t, \ell + 1, y-1} \[s_{\ell+1, y-1 ,t}\]$ is a summand of the $\ket{\o 2}$ codeword. 

Now, we observe that
\bea
\bra{\o 1}\lambda_1 (c_{\ell,y,t}\[s_{\ell,y,t}\]) &= \o{c_{y+1, t,\ell-1}}\cdot d_1\cdot \abs{\[s_{\ell -1, y+1,t}\]}^2\\
\bra{\o 2}\lambda_1 (c_{\ell,y,t}\[s_{\ell,y,t}\]) &= \o{c_{t, \ell + 1, y-1}}\cdot d_2 \cdot \abs{\[s_{t, \ell + 1, y-1}\]}^2
\eea
where $d_1 = c_{\ell, y ,t} (y + 1), d_2 = (\ell+1)c_{\ell,y, t}$. One easy way to force these quantities to be zero, is to state that $c_{y+1, t,\ell-1}, c_{t, \ell + 1, y-1} = 0$. This condition is true if and only if $\[s_{y + 1, t \ell - 1}\], \[s_{t, \ell + 1, y-1}\] \not\in S_{\ket{\o 0}}$. In other words, we do not want to include these vectors in the $\ket{\o 0}$ codeword.

What we have just demonstrated is that if we have some $\[s_{\ell, y ,t}\] \in S_{\ket{\o 0}}$, then $\[s_{y+1, t, \ell -1}\], \[s_{t, \ell + 1, y-1}\] \not\in S_{\ket{\o 0}}$ implies
\bea
\bra{\o 1}\lambda_1(c_{\ell, y ,t} \[s_{\ell, y ,t}\]) = \bra{\o 2}\lambda_1(c_{\ell, y ,t} \[s_{\ell, y ,t}\]) = 0.
\eea
In this case, $\lambda_1$ introduced 2 symmetric vectors that we want to avoid including as summands in the $\ket{\o 0}$ codeword. It turns out that $\lambda_4$ tells us that we want to avoid the same vectors that $\lambda_1$ tells us to avoid.

Given that $\[s_{\ell , y ,t}\] \in S_{\ket{\o 0}}$, $\lambda_1, \lambda_2, \lambda_3$ give us $6$ vectors total that should not be in $S_{\ket{\o 0}}$, namely
\bea\begin{array}{ccc}\label{eq: avoided_vectors}
\[s_{y+1, t ,\ell -1}\], & \[s_{y, t-1, \ell + 1}\] ,&\[s_{y-1, t+1, \ell}\],   \\
\[s_{t, \ell + 1, y-1}\], & \[s_{t+1, \ell - 1, y}\],  & \[s_{t-1, \ell, y + 1}\].
\end{array}
\eea
Note that the $i-$th column corresponds to the vectors that $\lambda_i$ tell us to avoid. To derive these, we would simply to the exact same procedure as before. We apply $\lambda_i$ to $\[s_{\ell, y ,t}\]$. The image vector will be the sum of two vectors. We then find which eigenspaces of the $Z$ operator the summands lie in. From there, we apply the appropriate power of the $X$ operator that would send the summands to the $+1$ eigenspace. The vectors in the $+1$ eigenspace are precisely the vectors that we do not want to include in $S_{\ket{\o 0}}$.

What we have just demonstrated is that we can methodically choose the vectors in $S_{\ket{\o 0}}$ such that
\bea
\bra{\o i} \lambda_j \ket{\o 0} = 0.
\eea
Using similar techniques as in Section \ref{reduction}, we find that
\bea
\bra{\o i}\lambda_j \ket{\o k} = 0,
\eea
for all $0 \leq i,k \leq 2$ and $1 \leq j \leq 6$.

While we can force the Knill-Laflamme conditions to be zero for error detection using the aforementioned method, we are not able to do this for error correction. For example, $\lambda_1^2$ acts in the following way
\bea
&\[ s_{\ell, y, t}\] \xrightarrow{\lambda_1}\\
&(\ell + 1)\[s_{\ell + 1, y-1, t} \]+ (y+1)\[s_{\ell-1, y+1, t}\] \xrightarrow{\lambda_1}\\
&(\ell + 2)(\ell + 1) \[s_{\ell + 2, y-2, t}\] + y(\ell + 1)\[s_{\ell, y, t}\]\\
& +\ell(y+1) \[ s_{\ell,y, t}\]\\
&+ (y+2)(y+1)\[s_{\ell - 2, y + 2 ,t}\].
\eea
We see that $\[s_{\ell, y, t}\]$ appears in the image of $\lambda_1^2$, therefore, $\[s_{\ell, y,t}\]^\dagger \lambda_1^2 \[s_{\ell, y, t}\] $ will never be 0 unless the coefficient of $\[s_{\ell, y, t}\]$ is 0. However, we can still force $\bra{\o i} \lambda_q \lambda_p \ket{\o j} = 0$.

For the following example, we will analyze the image of $\lambda_2\lambda_1$. We see that the action of $\lambda_2\lambda_1$ is given by
\bea
&\[ s_{\ell, y, t}\] \xrightarrow{\lambda_1}\\
&(\ell + 1)\[s_{\ell + 1,y-1, t} \] + (y+1)\[s_{\ell-1, y+1, t}\]
\xrightarrow{\lambda_2}\\
&(t+1)(\ell+1)\[s_{\ell, {y-1}, {t+1}}\]\\
&+ (\ell + 2)(\ell + 1) \[ s_{{\ell + 2}, {y-1}, {t-1}} \]\\
&+ \ell(y+1)\[s_{\ell, {y+1}, {t-1}}\]\\
&+ (t+1)(y+1)\[s_{{\ell - 2},{y+1}, {t + 1}}\].
\eea
Note that there are 4 summands in the image, namely
\bea\begin{array}{cc}
\[s_{\ell, {y-1}, {t+1}}\],& \[ s_{{\ell + 2}, {y-1}, {t-1}} \],\\
\[s_{\ell, {y+1}, {t-1}}\], & \[s_{{\ell - 2},{y+1}, {t + 1}}\].
\end{array}
\eea

Note that because $y \equiv t \mod 3$, we have that $y \pm 1 \equiv t \pm 1 \mod 3$ and thus $\[ s_{{\ell + 2}, {y-1}, {t-1}} \]$ and $\[s_{{\ell - 2}, {y+1}, {t + 1}}\]$ are in the $+1$ eigenspace of the $Z$ operator. Using similar logic as before, if $\[ s_{{\ell + 2}, {y-1}, {t-1}} \], \[s_{{\ell - 2}, {y+1}, {t + 1}}\] \not \in S_{\ket{\o 0}}$, in other words $c_{\ell + 2, y-1, t+1} = c_{\ell - 2, y + 1, t+1} = 0$, then it follows that
\bea
\bra{\o 0}\lambda_2 \lambda_1 (c_{\ell,y,t} \[s_{\ell, y, t}\]) = 0.
\eea
Next, we note that $\[s_{\ell, {y-1}, {t+1}}\]$ is a vector in the $\omega$ eigenspace of $Z$. From this, we assert that $X^{-1}\[s_{\ell, {y-1}, {t+1}}\] = \[s_{y-1, t+1, \ell}\] \not\in S_{\ket{\o 0}}$ so that
\bea
\bra{\o 1}\lambda_2 \lambda_1(c_{\ell, y, t}\[s_{\ell, y ,t}\]) = 0.
\eea
Finally, we note that $\[s_{\ell,{y+1},{t-1}}\]$ is in the $\omega^2$ eigenspace of the $Z$ operator. Therefore, we assert that $X^{-2}\[s_{\ell,{y+1},{t-1}}\] = \[s_{t-1, \ell, y+1}\] \not\in S_{\ket{\o 0}}$. Hence,
\bea
\bra{\o 2}\lambda_2\lambda_1 (c_{\ell, y, t}\[s_{\ell, y ,t}\]) = 0.
\eea
In conclusion, if we assert that $\[s_{\ell, y ,t}\] \in S_{\ket{\o 0}}$, then $\lambda_2\lambda_1$ tells us that we should avoid including the following vectors in $S_{\ket{\o 0}}$:
\bea\begin{array}{cc}
\[ s_{{\ell + 2}, {y-1}, {t-1}} \], & \[s_{{\ell - 2}, {y+1}, {t + 1}}\],\\
\[s_{\ell, {y-1}, {t+1}}\], & \[s_{t-1, \ell, y+1}\].
\end{array}
\eea
Notice that we already knew to avoid $\[s_{\ell, {y-1}, {t+1}}\], \[s_{t-1, \ell, y+1}\]$ because they are summands of the image of $\[s_{\ell, y ,t}\]$ under $\lambda_2$, which we handled previously. The new vectors that $\lambda_2\lambda_1$ tells us to avoid are the vectors that are two operations away from $\[s_{\ell, y ,t}\]$. 

Repeating a similar process for $\lambda_q\lambda_p$, we find that we should avoid including the following vectors in $S_{\ket{\o 0}}$:
\bean\begin{array}{ccc}\label{eq: avoided_vectors}
\[s_{y-1, t+1, \ell}\],& \[s_{y+1, t ,\ell -1}\], & \[s_{y, t-1, \ell + 1}\] \\
\[s_{t-1, \ell, y + 1}\],& \[s_{t, \ell + 1, y-1}\], & \[s_{t+1, \ell - 1, y}\] \\
\[s_{y-2,t, \ell + 2}\],& \[s_{y,t+2,\ell -2}\], & \[s_{y+2, t-2, \ell}\] \\
\[s_{t-2, \ell + 2, y}\],& \[s_{t+2, \ell, y-2}\], & \[s_{t, \ell - 2, y  +2}\]\\
\[s_{\ell-2, y + 1, t+1}\],& \[s_{\ell + 1, y + 1, t-2}\], & \[s_{\ell + 1, y-2, t+1}\].
\end{array}
\eean
Next, we define what it means for a codespace to be \textit{properly spaced}. Let $\[s_{\ell,y,t}\]$ be a member of the $+1$ eigenspace of the $Z$ operator. Define the set $A_{\ell, y ,t}$  as the following set
\bea
A_{\ell, y ,t} = \{\begin{array}{ccc}\label{eq: avoided_vectors}
\[s_{y-1, t+1, \ell}\],& \[s_{y+1, t ,\ell -1}\], & \[s_{y, t-1, \ell + 1}\] \\
\[s_{t-1, \ell, y + 1}\],& \[s_{t, \ell + 1, y-1}\], & \[s_{t+1, \ell - 1, y}\] \\
\[s_{y-2,t, \ell + 2}\],& \[s_{y,t+2,\ell -2}\], & \[s_{y+2, t-2, \ell}\] \\
\[s_{t-2, \ell + 2, y}\],& \[s_{t+2, \ell, y-2}\], & \[s_{t, \ell - 2, y  +2}\]\\
\[s_{\ell-2, y + 1, t+1}\],& \[s_{\ell + 1, y + 1, t-2}\], & \[s_{\ell + 1, y-2, t+1}\]
\end{array}\}
\eea
We say that a codespace is properly spaced if for all $\[s_{\ell, y ,t}\] \in S_{\ket{\o 0}}$, $S_{\ket{\o 0}} \cap A_{\ell, y ,t} = \emptyset$. One consequence of a codespace being properly spaced is that it follows that the following conditions are satisfied
\bea
\bra{\o i} \lambda_k \ket{\o j} &= 0\\
\bra{\o i} \lambda_q \lambda_p \ket{\o j} &= 0,
\eea
for $p \not= q$,$0 \leq i, j \leq 2$, and $1 \leq j \leq 6$.

When working in large enough representations, we have more than enough degrees of freedom to concoct properly spaced codespaces. For example, consider the codespace in the $(37,0)$ irrep given by
\bea
\ket{\o 0} &= a_0 \[s_{35,1,1} \] + a_1 \[s_{11,1,25}\] \\
&+ b_1 \[s_{11,25,1} \] + c_0 \[s_{9,14,14}\]\\
\ket{\o 1} = X\ket{\o 0} &= a_0 \[s_{1,35,1}\] + a_1 \[s_{25, 11, 1}\]\\
&+ b_1 \[s_{1, 11, 25} \] + c_0 \[s_{14, 9,14}\]\\
\ket{\o 2} = X\ket{\o 1} &= a_0 \[s_{1,1,35} \] + a_1 \[s_{1,25,11}\]\\
&+ b_1 \[s_{{25}, 1, {11}} \]+ c_0 \[s_{14, 14, 9}\].
\eea
We observe that $\lambda_2 \lambda_1 \[s_{35, 1, 1}\]$ given by
\bea
\lambda_2 \lambda_1 \[s_{35,1,1} \] &= 2\cdot 36 \[s_{35, 0 ,2}\] + 37\cdot 36 \[s_{37, 0 ,0 }\]\\
&+ 35 \cdot 2 \[s_{{35}, 2, 0 }\] + 2\cdot 2 \[s_{{34}, 2, 2}\].
\eea
By construction, one can check that $\[s_{35, 0,2}\] \not\in S_{\ket{\o 1}}$, $\[s_{37,0,0}\], \[s_{34,2,2}\], \not\in S_{\ket{\o 0}}$, and $\[s_{35,2,0}\] \not\in S_{\ket{\o 2}}$, and hence
\bea
\bra{\o i} \lambda_2\lambda_1 a_{0}\[s_{35,1,1}\] = 0,
\eea
for $i = 0 ,1, 2$. The reader can check that this code detects $\lambda_i$ errors, and we will give a specific code in Section \ref{sec: C-(37,0)} that has the same $S_{\ket{\o 0}}$ and corrects $\lambda_i$ errors.

In conclusion, we can satisfy most of the $\lambda_i$ Knill-Laflamme conditions by constructing a properly spaced codespace. Once we have a properly spaced codespace, the only nontrivial constraints introduced by the $\lambda_i$ operators are introduced by $\bra{\o i} \lambda_p^2 \ket{\o j}$.

\subsection{Satisfying $H_i$ Constraints}

The $H_i$ constraints play a huge role in determining whether or not our system of equations can be solved. First off, we note that the $H_i$ operators preserve the eigenspaces of the $Z$ operator. Therefore, if we have a codespace that satisfies $\bra{\o i} \lambda_q \ket{\o j} = 0$, then $H_p \lambda_q\ket{\o j} = c_2 \lambda_q \ket{\o j}$, and thus $\bra{\o i} H_p\lambda_q \ket{\o j} = c_2 \cdot 0 = 0$. Consequently, if we want to find a code that corrects $\lambda_i$ and $H_j$ errors, as long as the $\lambda_i$ conditions are satisfied, we just need to worry about the $\bra{\o i} H_k \ket{\o j}$ and $\bra{\o i} H_k^2 \ket{\o j}$ constraints.

To apply some clever tricks to simplifying the $H_i$ constraints, it's helpful to understand some special relationships between vectors. One special relationship exists between the vectors $\[s_{\ell, y, t}\]$ and $\[s_{\ell ,t, y}\]$, where $y \equiv t \mod 3$. Denote by $v$ the symmetric vector given by
\bea
v = X^2 \(c_{\ell, y ,t }\[s_{\ell, y ,t}\] + c_{\ell , t ,y} \[s_{\ell, t, y}\]\).
\eea
Observe the following
\bea
&v^\dagger H_1 v = p \cdot(y-t)^2 (\abs{c_{\ell, y ,t}}^2 - \abs{c_{\ell, t, y}}^2),
\eea
where $p =\frac{(\ell + y + t)!}{\ell! y! t!}$. Therefore, if 
\bea
\abs{c_{\ell, y ,t}}^2 - \abs{c_{\ell, t ,y}}^2 = 0,
\eea 
this would imply that $\abs{c_{\ell, y ,t}} = \pm \abs{c_{\ell, t ,y}}$. One simplification that we make is the simplification that $c_{\ell, y ,t} = c_{\ell, t ,y}$. In other words, if we include $\[s_{\ell, y ,t}\]$ in the codeword $\ket{\o 0}$, we want to include $\[s_{\ell ,t, y}\]$ in $\ket{\o 0}$ as well because it makes solving the equations easier. In general, $\ket{\o 0}$ may contain the following superposition of basis vectors
\bea
\sum_{i} (\alpha_i \[s_{{\ell_i},{t_i}, {y_i}}\] + \beta_i \[s_{{\ell_i}, {y_i} , {t_i}}\]),
\eea
and the following contribution of these vectors to $\bra{\o 2} H_1 \ket{\o 2}$ is given by
\bea
\sum_i p_i (y_i - t_i)^2 \(\abs{\alpha_i}^2 - \abs{\beta_i}^2\),
\eea
where $p_i = \frac{\(\ell_i + t_i + y_i\)!}{\ell_i! y_i! t_i!}$. As we can see, setting $\alpha_i = \beta_i$ in most cases will make $\bra{\o 2} H_1 \ket{\o 2} = 0$. We should note that we want to make sure that $\[s_{\ell ,y ,t}\]$ is not within 2 operations of $\[s_{\ell, t, y}\]$ because then the $\lambda_k$ conditions may no longer be satisfied.

The next type of special vector is the vector of the type $\[ s_{\ell, y, y}\]$. This is because
\bea
H_1 \(X^2(\alpha \[s_{\ell, y, y}\])\) = 0.
\eea
Hence, these vectors have no contribution in the equation $\bra{\o 2}H_1\ket{\o 2} = C_{H_1}$. However, the contributions that these vectors make to the other equations allow for us to manipulate the system of equations so that the system has a solution.

\subsection{Finding a Code in $(37,0)$}\label{sec: C-(37,0)}

Next, we will go through the process of finding an error correction code in the $(37,0)$ irrep of $SU(3)$.

First, we start by presenting the following properly spaced codespace,
\bea
\ket{\o 0} &= a_0 \[s_{35, 1,1} \] + a_1 \[s_{11,1,25}\] \\
&+ b_1 \[s_{11, 25, 1}\]
+ c_0 \[s_{9,14, 14}\].
\eea
It follows that
\bea
\bra{\o i} \lambda_q \lambda_p \ket{\o j} &= 0\\
\bra{\o i} \lambda_q \ket{\o j} &= 0
\eea
for $q \not= p$. From the previous section, we also know that
\bea
\bra{\o i} H_q \lambda_p \ket{\o j} = 0.
\eea
Furthermore, we know that
\bea
\bra{\o i} H_1 \ket{\o j} = 0,
\eea
for $i \not = j$ because the codewords are orthogonal to each other.

Next, we analyze the constraints that are not satisfied trivially. We will analyze the constraints introduced by $H_i$ errors first. We begin by observing the following
\bea
\bra{\o 0} H_1 \ket{\o 0} &= 34p_1\abs{a_0}^2 + 10 p_2\abs{a_1}^2 \\
& - 14p_2\abs{b_1}^2 - 5p_3\abs{c_0}^2\\
 \bra{\o 1} H_1 \ket{\o 1} &= -34p_1\abs{a_0}^2 +14p_2\abs{a_1}^2 \\
 & - 10p_2\abs{b_1}^2 + 5p_3\abs{c_0}^2\\
 \bra{\o 2} H_1 \ket{\o 2} &= 34p_2\abs{b_1}^2 - 34p_2\abs{a_1}^2.
\eea
where $p_i$ are the dot product multiplicities and
\bea\begin{array}{ccc}
p_1 = \frac{37!}{35!}, & p_2 = \frac{37!}{25! \cdot 11!}, & p_3 = \frac{37!}{14!\cdot 14! \cdot 9!}.
\end{array}
\eea
As mentioned earlier, we can make the assumption that $\abs{b_1} = \abs{a_1} $ to make the equations simpler. When doing this we get,
\bea
-\bra{\o 1} H_1 \ket{\o 1} &= \bra{\o 0} H_1 \ket{\o 0}\\
&= 34p_1\abs{a_0}^2 -4p_2\abs{a_1}^2\\
&- 5p_3\abs{c_0}^2\\
&=0\\
\bra{\o 2} H_1 \ket{\o 2} &= 0.
\eea
One thing to note here is that many times, in this step, we get a quadratic equation similar to the one in the above equations. However, we can determine whether or not the system is solvable if there are a mix of positive and negative coefficients. For instance, if we would have chosen $\ket{\o 0}$ to be
\bea
\ket{\o 0} &= p \[s_{35, 1 ,1} \] +  q_1\[s_{23, 1 ,13}\]\\
&+ q_1 \[s_{23, 13, 1} \]
+ r \[s_{17, 10 ,10}\],
\eea
then we would have had the following equation
\bea
\bra{\o 0} H_1 \ket{\o 0} &= 34p_1 \abs{p}^2 + 32c_1 \abs{q_1}^2\\
&+ 7 c_2 \abs{r}^2\\
&= 0,
\eea
with $c_1, c_2 > 0$. As one can see, this equation has no non-trivial solution. Therefore, we want to choose our codespace such that in $\ket{\o 0}$, the tensor power of $0$ is greater than the tensor power of $1$ in some basis vectors and less than the tensor power of $1$ in other basis vectors. This will ensure that $\bra{\o 0} H_1 \ket{\o 0} = 0$ has nontrivial solutions.

Next, we compute the second order $H_i$ constraints. Note that we did not compute $\bra{\o i} H_2 \ket{\o j}$ because we are using the assumption that if the $H_1$ conditions for error detection are satisfied, then the $H_2$ conditions will be satisfied as well. For error correction, however, we have to check the second order $H_2$ conditions as well. We begin by observing the following
\bea
\bra{\o 0} H_1^2 \ket{\o 0}&= 34^2p_1\abs{a_0}^2 +10^2 p_2\abs{a_1}^2 \\&+14^2p_2\abs{b_1}^2
+25p_3\abs{c_0}^2 \\
\bra{\o 1} H_1^2 \ket{\o 1} &= 34^2p_1\abs{a_0}^2 +10^2 p_2\abs{a_1}^2 \\& +14^2p_2\abs{b_1}^2
+25p_3\abs{c_0}^2\\
\bra{\o 2} H_1^2 \ket{\o 2} &= 34^2 p_2\abs{b_1}^2 + 34^2 p_2\abs{a_1}^2.
\eea
Next, we use the assumption that $\abs{a_1} = \abs{b_1}$, to conclude that
\bea
\bra{\o 1} H_1^2 \ket{\o 1} &= \bra{\o 0} H_1^2 \ket{\o 0}\\
&= 34^2p_1\abs{a_0}^2 +25p_3\abs{c_0}^2\\
& + (10^2 + 14^2) p_2\abs{a_1}^2\\
\bra{\o 2} H_1^2 \ket{\o 2} &= 2\cdot 34^2 p_2\abs{a_1}^2.
\eea
The error correction conditions tells us that all the quantities must be the same, therefore, it follows that
\bea
&34^2p_1\abs{a_0}^2 +25p_3\abs{c_0}^2\\
&+ (10^2 + 14^2) p_2\abs{a_1}^2\\
& = 2\cdot 34^2 p_2\abs{a_1}^2 
\eea
Note again that because $10^2 + 14^2 < 2\cdot 34^2$, the equation has a chance of having nontrivial solutions.

Next, we check the $H_1H_2$ conditions. Note that $H_1 H_2 = H_2 H_1$, therefore, we only need to check $\bra{\o i} H_1H_2 \ket{\o j} = C$. The process is similar to the other processes, so we get the following
\bea
\bra{\o 2} H_1 H_2 \ket{\o 2} &= \bra{\o 0} H_1 H_2 \ket{\o 0}\\
&= -(576) p_2\abs{a_1}^2\\
\bra{\o 1} H_1H_2 \ket{\o 1} &= -34^2 p_1\abs{a_0}^2 \\
&+ 280p_2\abs{a_1}^2- 5^2 p_3\abs{c_0}^2,
\eea
and hence
\bea
&-34^2 p_1\abs{a_0}^2
- 5^2p_3\abs{c_0}^2
 +280p_2\abs{a_1}^2\\
 &= -(576) p_2\abs{a_1}^2.
\eea
By construction, we see that this equation will also have non-trivial solutions because there is a mixture of positive and negative coefficients. In summary, the $H_i$ constraints for error correction are given by
\bea
&34p_1\abs{a_0}^2 -4p_2\abs{a_1}^2 - 5p_3\abs{c_0}^2 = 0\\
&34^2p_1\abs{a_0}^2 + 296 p_2\abs{a_1}^2 +25p_3\abs{c_0}^2
\\& = 2\cdot 34^2 p_2\abs{a_1}^2\\
&-34^2 p_1\abs{a_0}^2 + 280p_2\abs{a_1}^2 - 25 p_3\abs{c_0}^2\\
&= -(576) p_2\abs{a_1}^2.
\eea

Next, we compute the non-trivial $\lambda_i$ conditions. Note that $X^\dagger \lambda_1^2 X = \lambda_2^2$. Therefore, using the same logic as before, we only need to check $\bra{\o i} \lambda_1^2 \ket{\o i}$. The equations that we get for $\lambda_1^2$ are
\bea
\bra{\o 1} \lambda_1^2 \ket{\o 1} &= \bra{\o 0} \lambda_1^2 \ket{\o 0}\\
&= (106)p_1\abs{a_0}^2\\
&+ (620)p_2\abs{a_1}^2\\
&+ (275) p_3\abs{c_0}^2\\
\bra{\o 2} \lambda_1^2 \ket{\o 2} &= 4p_1\abs{a_0}^2 + 152p_2\abs{a_1}^2\\
&+ 420 p_3\abs{c_0}^2,
\eea
thus, we have the following result
\bea
&(106)p_1\abs{a_0}^2 + (620)p_2\abs{a_1}^2\\
&+ (275) p_3\abs{c_0}^2\\
&= 4p_1\abs{a_0}^2 + 152p_2\abs{a_1}^2 \\
&+ 420 p_3\abs{c_0}^2.
\eea
Putting everything together, we get
\bea
&34p_1\abs{a_0}^2 -4p_2\abs{a_1}^2 - 5p_3\abs{c_0}^2\\
&= 0\\
&1156p_1\abs{a_0}^2 + 296 p_2\abs{a_1}^2\\
&+25p_3\abs{c_0}^2\\
&= 2312p_2\abs{a_1}^2-1156 p_1\abs{a_0}^2
\\
&+ 280p_2\abs{a_1}^2
- 25p_3\abs{c_0}^2\\
&= -576 p_2\abs{a_1}^2\\
&106p_1\abs{a_0}^2 + 359p_2\abs{a_1}^2 \\
&+ 275 p_3\abs{c_0}^2\\
&= 4p_1\abs{a_0}^2 + 152p_2\abs{a_1}^2 \\
&+ 420 p_3\abs{c_0}^2.
\eea
A solution to the system of equations yields the following codewords,
\bea
\ket{\o 0} &= \frac{1}{d}\(\array{c} \sqrt{670371601625} \[s_{35,1,1} \]\\ + \sqrt{\frac{10214875}{168}} \[s_{11,1,25}\]\\ + \sqrt{\frac{10214875}{168}} \[s_{11,25,1} \]
\\+ \[s_{9,14,14}\] \endarray \)\\
\ket{\o 1} &= \frac{1}{d}\(\array{c} \sqrt{670371601625} \[s_{1,35,1}\]\\ + \sqrt{\frac{10214875}{168}} \[s_{25,11,1}\]\\ + \sqrt{\frac{10214875}{168}} \[s_{1,11,25}\]
\\+ \[s_{14,9,14}\] \endarray \)\\
\ket{\o 2} &= \frac{1}{d}\(\array{c} \sqrt{670371601625} \[s_{1,1,35} \]\\ + \sqrt{\frac{10214875}{168}} \[s_{1,25,11}\]\\ 
+ \sqrt{\frac{10214875}{168}} \[s_{25,1,11} \]
\\+ \[s_{14,14,9}\] \endarray \),
\eea
where $d = \sqrt{8586854127423000}$.
\clearpage

\section{Conclusion and Future Work}
\label{conclusions}

We have found that there exists an error correction code in the $(37,0)$ irrep of $SU(3)$; however, there is reason to believe that this is not the smallest possible representation that has an error correction code. One possible extension of this project is to find a lower bound on the dimension of the space that contains an error correcting code. Another possible extension is to find a recipe for generating error correction codes in $SU(n)$ for $n \geq 4$. $SU(3)$ just so happend to be fairly similar to $SU(2)$, but as $n$ increases, things may potentially become more complicated.

A natural question to ask is whether $\SU(3)$ representations correspond to ``physical'' state spaces.
Harmonic oscillators, molecules, and large nuclear spins all give relevant examples of $\SL(2,\mathbf{R})$, $\SO(3)$, and $\SU(2)$ representations.
Within physics, $\SU(3)$ is famous for its role in the standard model, but we're hardly advocating for building a quantum computer out of individually addressable quarks!
However, one can look to the Schwinger representation to provide a motivating physical instantiation.
This useful representation of $\SU(2)$~\cite{schwinger_angular_1952}, which uses number-preserving operations on a collection of harmonic oscillators, also provides a means of realizing $\SU(3)$ symmetry via linear optics (beamsplitters and phase shifters)~\cite{chaturvedi_schwinger_2002}.  Encoding logical qutrits can also yield computational advantages~\cite{gokhale2019asymptotic}.

Designing codes using $\SU(3)$ is also a natural step towards building a general prescription for constructing quantum codes from higher dimensional unitary group representations, showcasing some of the complexities hidden in the special case of $\SU(2)$ \cite{gross_designing_2021}.
The result of our labor is the discovery of new codes embedding three-dimensional logical quantum systems in physical state spaces given by $\SU(3)$ irreps.

\section{Acknowledgments}

We thank Dripto Debroy for providing comments on the manuscript.
We thank the summer internship program at Google Quantum AI for hosting this research.
\appendix

\section{Structural Proof of $(p,0)$ Irrep}

First, we will show that the action of the $Y_1$ operator preserves symmetric vectors. We wish to show that
\bea
\[0^\ell \otimes 1^y \otimes 2^t \] \xrightarrow{Y_1} (y+1) \[0^{\ell-1} \otimes 1^{y+1} \otimes 2^t \].
\eea

To show this, we will first show that any permutation of the vector $\ket{0^{\ell - 1} \otimes 1^{y+1} \otimes 2^t}$ is the image of some permutation of $\ket{0^\ell \otimes 1^y \otimes 2^t}$ under $Y_1$. For the rest of this section define $\ket{s_{\ell, y, t}}$ as the following
\bea
\ket{s_{\ell, y ,t}} = \ket{0^\ell \otimes 1^y \otimes 2^t}.
\eea
Given any permutation $\ket{r}$ of $\ket{s_{{\ell - 1},{y+1},t}}$, we know that there exists some permutation $\sigma$ such that
\bea
\ket{r} = \sigma \ket{s_{{\ell - 1} ,{y+1},t}}.
\eea
Hence, we see that $\sigma\ket{s_{{\ell}{y},t}}$ will have nearly the same tensor product decomposition as $\ket{r}$, the only difference is that $\ket{r}$ has a 1 in a spot where $\sigma\ket{s_{{\ell}, {y}, t}}$ has a 0. Suppose $\sigma\ket{s_{{\ell}, {y}, t}}$ differs from $\ket r$ at the $i$th tensor product. Therefore, we see that
\bea
\(I^{\otimes i -1} \otimes Y_1 \otimes I^{\otimes n - i }\)\sigma\ket{s_{{\ell}, {y}, t}} = \ket{r},
\eea
where $n = k + \ell + t$. We have shown that there exists some permutation of $\sigma\ket{s_{{\ell}, {y}, t}}$ that maps to $\ket{r}$. In other words, there exists some vector in the class $\[0^{\ell} \otimes 1^{y} \otimes 2^t\]$ that maps to $\ket{r}$. Hence, every vector in the class $\[0^{\ell - 1} \otimes 1^{y + 1} \otimes 2^t\]$ is reached. A similar argument can be made for $Y_2$ to show that
\bea
\[0^\ell \otimes 1^y \otimes 2^t\] \xrightarrow{Y_2} (t+1)\[ 0^\ell \otimes 1^{y-1} \otimes 2^{t+1}\]
\eea

To see why the multiplicity is $y+1$, we begin by denoting the transposition $\tau_{(i,j)}$ as the permutation that swaps the $i$th and $j$th letter. We claim that
\bea
&\(I^{\otimes \ell + j -1} \otimes Y_1 \otimes I^{n-y}\)\tau_{(\ell, \ell + j)}\ket{s_{{\ell}, {y}, t}}\\
&= 
\(I^{\otimes \ell - 1}\otimes Y_1 \otimes  I^{n-\ell}\)\ket{s_{{\ell}, {y}, t}},
\eea
for $0 \leq j \leq y$. Indeed, we have
\bea
&\(I^{\otimes \ell + j-1} \otimes Y_1 \otimes I^{\otimes (n-y)}\)\tau_{(\ell, \ell + j)}\ket{s_{{\ell}, {y}, t}}\\
&= \(I^{\otimes \ell + j-1} \otimes Y_1 \otimes I^{\otimes (n-y)}\)\ket{w}\\
&= \ket{s_{{\ell - 1}, {y + 1}, t}}\\
&= \(I^{\otimes \ell - 1}\otimes Y_1 \otimes  I^{n-\ell}\)\ket{s_{{\ell}, {y}, t}},
\eea
where $\ket{w}$ is the following vector
\bea
\ket{w} = \ket{0^{\ell-1} \otimes 
1^{j} \otimes 0 \otimes  1^{y -j} \otimes 2^t}.
\eea

Because there are $y+1$ different transpositions that map to the same image, we see that the image multiplier is $y+1$, as desired.

As a consequence of the above results, we have that for $y \equiv t \mod 3$,
\bea
&\[0^\ell \otimes 1^y \otimes 2^t\] \xrightarrow{Y_1Y_2Y_1}\\
&(y+1)(t+1)(y+1)\[0^{\ell -2} \otimes 1^{y+1} \otimes 2^{t+1} \],
\eea
therefore, it follows that $\[0^{\ell - 2} \otimes 1^{y+1} \otimes 2^{t+1}\]$ is a member of the irrep. Consequentially, if we start of with $\[0^n\]$, we can repeatedly apply the family of operators $Y_1^{3r}, (Y_1Y_2Y_1)^w,$ and $Y_2^{3q}$ to generate all of the classes of vectors that correspond to the $+1$ eigenspace of the $Z$ operator.

We have shown that the symmetric vectors are a subspace of the irrep, now it is time to show that there are $\frac{1}{2}(p+1)(p+2)$ symmetric vectors. To prove this, the problem reduces down to showing that there are $\frac{1}{2}(p+1)(p+2)$ ways of summing 3 non-negative numbers to $p$.

First, we show that there are $p+1$ ways of summing up to $p$ using 2 non-negative numbers. Note that $p \geq p- x_0 \geq 0$ for $0 \leq x_0 \leq p$. Therefore, there are $p+1$ $x_0$ that satisfy the inequality, which proves that there are $p+1$ ways to sum up two non-negative integers to $p$. Next, if we want to find the number of ways to sum of $3$ non-negative integers to $p$, we can do the following. Note note that there are $p+1$ solutions to $p \geq p - 0 + x_0 + x_1 \geq 0$. Similarly, there are $p$ solutions to $p - 1 \geq (p-1) + x_0 + x_1 \geq 0$. In general, there are $p + 1 - q$ solutions to
\bea
x_0 + x_1 = p-q.
\eea
Therefore, if we let $q$ range from $0$ to $p$, this gives us the number of ways to sum up 3 numbers to $p$. Consequently, summing up the number of solutions, we have
\bea
\sum_{j=0}^p (p+1 - j) &= \sum_{i=1}^{p+1} i\\
&= \frac{1}{2}(p+1)(p+2).
\eea
Hence, we know that there exists a bijection between solutions of the equation
\bea
\ell + y + t = p,
\eea
and symmetric vectors
\bea
\[0^\ell \otimes 1^y \otimes 2^t\],
\eea
which proves that the symmetric vectors are a basis for the $(p,0)$ irrep.

\bibliography{references}
\end{document}